\documentclass[superscriptaddress,prb, reprint, longbibliography, 
 amsmath,amssymb,
 aps,
]{revtex4-2}
\usepackage{multirow}
\usepackage{graphicx}
\usepackage{dcolumn}
\usepackage{bm}
\usepackage{hyperref}
\hypersetup{colorlinks,allcolors=blue}
\begin{document}


\title{Exploring functionalized Zr$_2$N and Sc$_2$N MXenes as superconducting candidates with \textit{ab initio} calculations}
\author{Alpin N. Tatan}
\email{alpin.tatan@phys.s.u-tokyo.ac.jp}

\author{Osamu Sugino}

\affiliation{Department of Physics, Graduate School of Science, The University of Tokyo, 7-3-1 Hongo, Bunkyo-ku, Tokyo, 113-0033, Japan.}
\affiliation{The Institute for Solid State Physics, The University of Tokyo, 5-1-5 Kashiwanoha, Kashiwa, Chiba, 277-8581, Japan.}

\date{June 16, 2025}

\begin{abstract}
We study new superconductor candidates in functionalized MXenes Zr$_2$NS$_2$, Zr$_2$NCl$_2$, and Sc$_2$NCl$_2$ with \textit{ab initio} calculations based on density functional theory for superconductors (SCDFT). The superconducting transition temperature $(T_c)$ at ambient pressure is predicted to reach 9.48 K (Zr$_2$NS$_2$), with potential further improvements under applied strain. We note that the changes in the profiles of superconducting gap $(\Delta)$ and electron-phonon coupling $(\lambda)$ across the Fermi surface may be influenced by their modified electronic bandstructure components.
\end{abstract}

\maketitle

\section{\label{sec:level1}Introduction}
MXenes (M$_{n+1}$X$_n$) are two-dimensional (2D) materials made of two or more layers of transition metal (M) atoms sandwiching carbon or nitrogen (X) layers \cite{VahidMohammadi_MXenes}. MXenes have a hexagonal close-packed (hcp) crystal structure with a $P6_3/mmc$ space group symmetry, where the transition metals in M sites are close-packed, and the X atoms occupy octahedral sites in between the M atomic planes \cite{VahidMohammadi_MXenes}. This family of 2D materials generally has metallic, nonmagnetic structure with quite large density of states at Fermi level \cite{Khazaei_MXenes, Bekaert_MXene}, opening itself to many potential applications \cite{Khazaei_MXenes, VahidMohammadi_MXenes}. 

In its synthesis, MXenes are often obtained with additional surface terminations \cite{Khazaei_MXenes}. The presence of surface functional groups (T = Cl, S, $\dots$) may modify the properties of MXenes. In the case of Nb$_2$C, functionalization with Cl or S leads to superconductivity \cite{Kamysbayev,WANG2022101711, PAZNIAK2024100579, Xu_Nb2CT, Jiang_Angew, Sevik_2023,Sevik_corr} while the bare,  O- or F-terminated compounds do not superconduct at temperatures as low as 2 K \cite{Kamysbayev,WANG2022101711}. Mixture of surface terminations, e.g., Cl combined with F, O, etc. may also suppress superconductivity \cite{Xu_Nb2CT}. Although the nature of superconductivity in these compounds is unconfirmed, similarities to electron-phonon mechanism based on pressure dependence and specific heat measurements have been noted \cite{PAZNIAK2024100579}. \textit{Ab initio}  studies \cite{WANG2022101711, Sevik_2023, Sevik_corr} based on McMillan formula \cite{McMillan} have also reported comparable transition temperature $(T_c)$ values. 

For superconductivity research, the ability to design and tune the position of energy bands associated with desired electronic states, the position of phonon bands related to vibrational states, and consequently modify the electron-phonon coupling strength through their combination is particularly important. This tunability of electronic and phononic properties via surface functionalization offers a unique avenue for designing novel superconducting materials. As far as we know, superconductivity in functionalized MXenes remains largely unexplored: the Nb$_2$C family being the only confirmed system to exhibit this property. Historically, the presence of multiple systems sharing similar phenomena is beneficial for improving their comprehension, as lessons derived from one system may be utilized for others \cite{Dagotto2003}. The interrelated studies of various cuprate and nickelate superconductors exemplify this practice \cite{zhang2021nickelate,Ji_2021,Pickett1989,Nokelainen2020,furness2018, Tatan2022,zhang2020,Armitage2010,FOURNIER2015,Tatan_2024,Nomura_2022}. Hence, it is worthwhile to explore other MXene compounds whose superconductivity may be enabled with surface functionalization. While approaches using data science might be employed to reveal chemical trends, in this study, we adopt an approach of computationally searching for multiple stable materials with real phonon frequencies in a targeted subset of functionalized compounds, carefully investigating their physical properties using ab initio calculations, and presenting these results to experimentalists as insights into MXenes' superconducting properties. This is the main objective of this paper.

\begin{figure}[h]
    \centering
    \includegraphics[width = 1\linewidth]{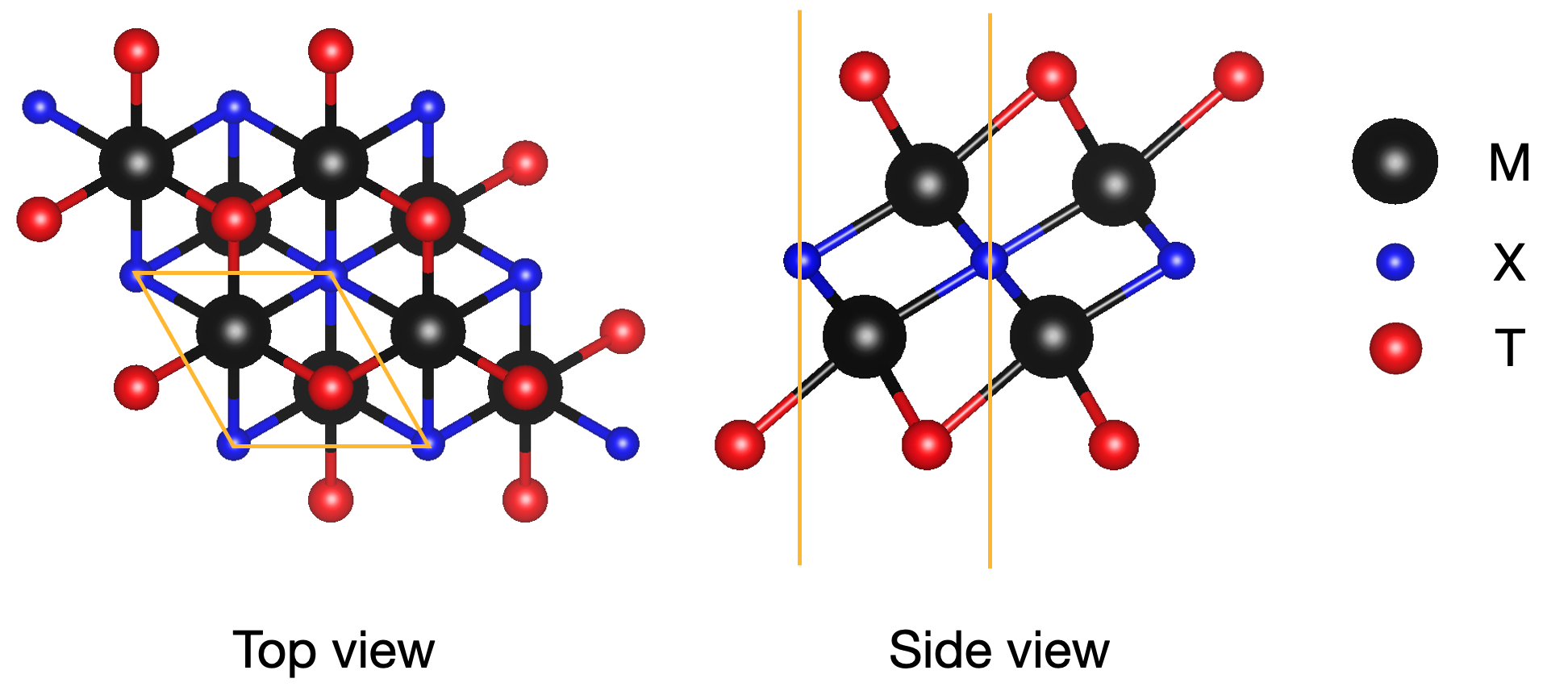}
    \caption{Top and side views of the crystal structure model for functionalized MXenes M$_2$XT$_2$. Solid orange lines mark the unit cell. Black, blue, and red spheres represent the transition metal atoms M, the C or N atoms X, and the functional group atoms T, respectively.}
    \label{fig:model}
\end{figure}

We leverage on past studies \cite{Qin_2020, lu2024first, Bekaert_MXene} to shortlist the vast combinations of MXenes and functional groups to be explored in this paper. We focus on the simplest case of $n=1$ as this group hosts an experimentally known superconductor (Nb$_2$CT$_2$). Among the possible configuration sites in the functionalized MXene M$_2$XT$_2$, the structure depicted in Figure \ref{fig:model} is generally deemed to be most stable \cite{Khazaei_MXenes, Sevik_2023, Sevik_corr, Bekaert_MXene_H, Qin_2020, lu2024first}. We consider functionalization with T = Cl and S, as these are known experimentally to induce superconductivity in Nb$_2$C \cite{Kamysbayev}. We remark that Nb$_2$CSe and Nb$_2$CNH were also reported to superconduct \cite{Kamysbayev}, but these cases may involve compounds containing high number of vacancies \cite{Sevik_2023} or with uncertain distribution of the T atoms over the MXenes that make their simulation nontrivial. Hence, we limit our scope only to explore the structures in Figure \ref{fig:model} with T = Cl and S in this paper.

A prior study \cite{Bekaert_MXene} has computationally screened the bare MXene compounds (M$_2$X) to look for potential superconductor candidates. Since we are interested specifically in achieving superconductivity via functionalization, we study compounds from Ref. \cite{Bekaert_MXene} which do not superconduct in their bare forms. We also leverage on prior stability screenings for functionalization with Cl and S \cite{Qin_2020, lu2024first}. From the candidates, we omit the non-metallic compounds (e.g., Sc$_2$CCl$_2$ \cite{Qin_2020}), and the experimentally synthesized ones with no reports on superconductivity (e.g., Ti$_2$CCl$_2$ or Zr$_2$CCl$_2$ \cite{Wang_Zr2CCl2}). This yields Zr$_2$NCl$_2$, Sc$_2$NCl$_2$ and Zr$_2$NS$_2$ as the remaining compounds that we select for this study. In particular, we will look for enhanced $T_c$ values in these functionalized compounds in contrast to their bare forms. For completeness, we have also attempted to compute Sc$_2$NS$_2$. However, we found this compound unstable with imaginary phonon frequencies, and thus it is not included in this paper. In addition, we will also demonstrate the possibility of strain-induced $T_c$ enhancement in the functionalized MXene compounds. For this purpose, we perform calculations on Zr$_2$NCl$_2$ and Zr$_2$NS$_2$ under several strain levels as a proof-of-concept exercise. We hope this demonstration may encourage a comprehensive straintronics study of MXenes in the future.

First-principles calculations based on density functional theory (DFT) and density functional perturbation theory (DFPT) are often used for searching new superconductor candidates. Combined with the phenomenological McMillan formula \cite{McMillan}, the $T_c$ is given by:
\begin{equation}
    T_c = \frac{\omega_\mathrm{ln}}{1.2} \mathrm{exp}\left[\frac{-1.04(1+\lambda)}{\lambda-\mu^{*}(1+0.62\lambda)}\right]
\end{equation}
where the electron-phonon coupling constant $\lambda$ and the averaged phonon frequency $\omega_\mathrm{ln}$ are obtainable from standard DFPT routine. The effect of screened Coulomb pseudopotential is parametrized by $\mu^{*}$, whose value is typically specified around $0.1-0.15$ \cite{Morel-Anderson,Sano2019}. However, the need for such empirical choice limits the theoretical predictive power and no superconducting gap information is accessible by the McMillan approach. Although it is possible to solve for the gap with the Migdal-Eliashberg (ME) formalism \cite{Lee_EPW,EPW_PhysRevB.87.024505}, this alternative method also requires the user to empirically specify the $\mu^{*}$ value in practice. Hence, we turn to the density functional theory for superconductors (SCDFT) \cite{Luders2005, Marques2005, Kawamura_benchmark, Sanna2020_PhysRevLett.125.057001, kawamura_ynibc_PhysRevB.95.054506} for its ability to solve the gap equation for $T_c$ without the need for empirical parameters. This SCDFT trait thus differentiates our work from other studies derived from the McMillan and Migdal-Eliashberg approaches.

This paper follows a similar strategy to preceding works \cite{Bekaert_MXene_H, Sevik_2023, Sevik_corr} in exploring possible superconductivity in functionalized MXenes. Section \ref{comp} outlines the computational methods. Section \ref{sect-results} computes the configuration in Figure \ref{fig:model} for $T_c$ of the target compounds. The $T_c$ trend is compared against electron-phonon coupling constant $\lambda$, electron density of states at Fermi level $N(E_\mathrm{F})$ (states/eV), and averaged Coulomb interaction $\mu_C$ to look for correlations. As with previous works \cite{Bekaert_MXene_H,Bekaert_MXene, Sevik_2023, Sevik_corr}, we shall find that $\lambda$ best correlates with $T_c$. The changes upon functionalization or applied strain are analyzed from the modified electronic bandstructure and phonon dispersion. The effects of ferromagnetic spin-fluctuations (SF) are also briefly discussed. Section \ref{sect-conclusions} concludes the paper.

\section{Computational Methods \label{comp}}
Normal-state DFT and DFPT calculations are carried out with \texttt{QUANTUM ESPRESSO} code \cite{QE1,QE2,Sohier_2017}. Exchange–correlation effects are treated with the generalized gradient approximation (GGA) using the Perdew – Burke – Ernzerhof (PBE) functional \cite{PBE_PhysRevLett.77.3865}. A vacuum of at least 15 \AA~ is included in the unit cell.  We use ultrasoft pseudopotentials from PSlibrary \cite{PSL_DALCORSO2014337}. The structure relaxation is performed with energy and force conversion thresholds of $10^{-6}$ and $10^{-5}$ in Ry atomic units, under 0.003675 Ry Gaussian smearing. The energy cutoff for wave function and charge density are 80 Ry and 640 Ry, respectively. SCDFT calculation is carried out with the \texttt{SUPERCONDUCTING-TOOLKIT} code \cite{Kawamura_benchmark}. For this procedure, the self-consistent charge density calculation uses the optimized tetrahedron method \cite{tetrahedron_PhysRevB.89.094515}. Spin-orbit interaction (SOI) effects are omitted, as they are expected to be small for Zr$_2$N and Sc$_2$N-based compounds \cite{Bekaert_MXene}. The $T_c$ is obtained by solving the SCDFT gap equation,
\begin{align}
        \Delta_{n\mathbf{k}} &= -\frac{1}{2}\sum_{n'\mathbf{k'}}\frac{K_{n\mathbf{k}n'\mathbf{k'}}(\xi_{n\mathbf{k}},\xi_{n'\mathbf{k'}})}{1+Z_{n\mathbf{k}}(\xi_{n\mathbf{k}})} \nonumber \\
        &\times \frac{ \Delta_{n'\mathbf{k'}} }{\sqrt{\xi^2_{n'\mathbf{k'}} + \Delta^2_{n'\mathbf{k'}} }} \tanh{\left(\frac{\sqrt{\xi^2_{n'\mathbf{k'}} + \Delta^2_{n'\mathbf{k'}} }}{2T}\right)}
\end{align} 
where $\xi_{n\mathbf{k}}$ is the Kohn-Sham eigenvalue measured from the Fermi level $E_\mathrm{F}$ at the band index $n$ and wavenumber $\mathbf{k}$. $T_c$ is the temperature where the superconducting gap $ \Delta_{n\mathbf{k}}$ vanishes at all $n$ and $\mathbf{k}$. $K_{n\mathbf{k}n'\mathbf{k'}}$ is an integration kernel made of interaction terms, for instance:
\begin{equation}
    K_{n\mathbf{k}n'\mathbf{k'}} = K^\mathrm{ep}_{n\mathbf{k}n'\mathbf{k'}} + K^\mathrm{ee}_{n\mathbf{k}n'\mathbf{k'}}
\end{equation}
consists of the electron-phonon $K^\mathrm{ep}_{n\mathbf{k}n'\mathbf{k'}}$ and the (screened) electron-electron interaction $K^\mathrm{ee}_{n\mathbf{k}n'\mathbf{k'}}$ terms. Similarly, $Z_{n\mathbf{k}}$ is a renormalization factor that arises from such interactions. Approximations to $K_{n\mathbf{k}n'\mathbf{k'}}$ and $Z_{n\mathbf{k}}$ have been derived elsewhere \cite{Luders2005,Marques2005,Kawamura_benchmark,nomoto_PhysRevB.101.014505,Sanna2020_PhysRevLett.125.057001}. For the electron-phonon kernels, we use the expressions of Ref. \cite{Sanna2020_PhysRevLett.125.057001}. The random phase approximation (RPA)-derived electron-electron interaction kernels are computed with at least 20 empty bands, following Ref. \cite{Kawamura_benchmark}.

\begin{table}[h]
\caption{The $\mathbf{k}$- and $\mathbf{q}$-point grids for SCDFT procedure.}
\label{tbl:grid}
\begin{ruledtabular}
\begin{tabular}{lc}
Calculation steps  & $\mathbf{k}$- and $\mathbf{q}$-point grids \\
\hline
self-consistent charge density & $\mathbf{k}$: medium grid   \\
phonons (DFPT) & $\mathbf{k}$: medium; $\mathbf{q}$: coarse\\
electron-phonon coupling & $\mathbf{k}$: coarse; $\mathbf{q}$: coarse  \\
Energy dispersion (one-shot) & $\mathbf{k}$: fine grid \\
Kohn-Sham orbitals (one-shot) & twin $\mathbf{k}$-point grids\\
screened Coulomb interaction & $\mathbf{q}$: coarse grid \\
SCDFT gap equation & $\mathbf{k}$: coarse grid \\
\end{tabular}
\end{ruledtabular}
\end{table}

The information about the $\mathbf{k}$- and $\mathbf{q}$-point grids for each SCDFT procedure step \cite{Kawamura_benchmark} is summarized in Table \ref{tbl:grid}. For Nb$_2$C systems, Sc$_2$NCl$_2$ and Zr$_2$NCl$_2$, $8 \times 8 \times 1$, $16 \times 16 \times 1$, and $32 \times 32 \times 1$ are used as the coarse, medium, and fine grid densities. 128 $\mathbf{k}$-points are used for the twin grids. For Zr$_2$N and Zr$_2$NS$_2$, the coarse, medium, and fine grid densities are increased to $10 \times 10 \times 1$, $20 \times 20 \times 1$, and $40 \times 40 \times 1$ with twin grids of 200 $\mathbf{k}$-points to improve convergence. We generally follow the study of elemental superconductors \cite{Kawamura_benchmark}, in which a satisfactory convergence of the results was demonstrated with similar parameters.

\section{Results and Discussions \label{sect-results}}
\subsection{$T_c$ of Nb$_2$C, Nb$_2$CCl$_2$, and Nb$_2$CS$_2$}
First, we briefly check our approach against experiment \cite{Kamysbayev} and McMillan formula calculations  \cite{Bekaert_MXene,Sevik_2023, Sevik_corr} for Nb$_2$C, Nb$_2$CCl$_2$, and Nb$_2$CS$_2$ in Table \ref{tbl:bench}. Our SCDFT-computed $T_c$ values are comparable with those preceding results. The relative $T_c$ trend, i.e., $T_c$ (Nb$_2$C) $<$ $T_c$ (Nb$_2$CCl$_2$) $\lesssim$ $T_c$ (Nb$_2$CS$_2$) is also well-captured. This also supports the conjectured phonon-mediated mechanism in these materials \cite{PAZNIAK2024100579, Sevik_2023,Sevik_corr, Bekaert_MXene_H, Bekaert_MXene}.

\begin{table}[h]
  \caption{The superconducting transition temperature $T_c$ (K) of Nb$_2$C, Nb$_2$CCl$_2$ and Nb$_2$CS$_2$.}
  \label{tbl:bench}
\begin{ruledtabular}
\begin{tabular}{lccc}
    $T_c$ (K)  & Experiment & DFT-McMillan & SCDFT  \\
    & \cite{Kamysbayev} & \cite{Bekaert_MXene,Sevik_2023,Sevik_corr} & (this work) \\
    \hline
Nb$_2$C & less than 2 K & less than 1 K & less than 2 K \\
Nb$_2$CCl$_2$ & 6.0 & 9.6 & 8.2 \\
Nb$_2$CS$_2$ & 6.4 & 10.7 & 8.4 \\
\end{tabular}
\end{ruledtabular}
\end{table}

Having confirmed the applicability of SCDFT for functionalized Nb$_2$C, we use it next for exploring Zr$_2$N and Sc$_2$N-based materials. 

\subsection[Functionalized zirconium-nitride and scandium-nitride MXene compounds]{Functionalized Sc$_2$N and Zr$_2$N compounds}
\subsubsection{Overview}

The computed values for $T_c$, $\lambda$, $N(E_\mathrm{F})$, and $\mu_C$ for the MXenes studied in this paper are provided in Table \ref{tbl:all}. 

\begin{table}[h]
  \caption{The superconducting transition temperature $T_c$ (K), electron-phonon coupling constant $\lambda$, electron density of states at Fermi level $N(E_\mathrm{F})$ (state/eV), and averaged Coulomb interaction $\mu_C$ for functionalized MXene compounds studied in this paper.}
  \label{tbl:all}
\begin{ruledtabular}
\begin{tabular}{lcccccc}
    Compounds & $T_c$   & $\lambda$ & $N(E_\mathrm{F})$  & $\mu_C$ \\
    \hline
    Sc$_2$NCl$_2$  & 1.79  & 0.31 & 1.44 & 0.27 \\
    Zr$_2$NCl$_2$ & 2.16  & 0.33 & 1.18 & 0.20 \\
    Zr$_2$NS$_2$ & 9.48 & 0.60 & 1.25 & 0.19 \\
\end{tabular}
\end{ruledtabular}
\end{table} 

Note that the empirical parameter $\mu^{*}$ is related to $\mu_C$:

\begin{equation}
    \mu_C = N(E_\mathrm{F}) \left\langle K^{ee}_{n\mathbf{k}n'\mathbf{k}'} \right\rangle \label{mu_C}
\end{equation}
\begin{equation}
     \mu^{*} = \frac{\mu_C}{1+\mu_C \ln \frac{E_c}{\omega_c}}
\end{equation}
where $ \left\langle\right\rangle$ indicates a Fermi surface average. $E_c$ and $\omega_c$ are electronic and phononic cutoff energies, whose values are often arbitrarily specified  \cite{Simonato_PhysRevB.108.064513, Akashi_2012_PhysRevB.86.054513}. This ambiguity makes it less meaningful to compare $\mu^{*}$ computed in this way against the customary values of 0.1 - 0.15 in the McMillan formalism \cite{kawamura_ynibc_PhysRevB.95.054506,Akashi_2012_PhysRevB.86.054513}. Instead, comparing the values of $\mu_C$ is preferred \cite{Akashi_2012_PhysRevB.86.054513}. We find our $\mu_C$ values to be comparable to those of elemental superconductors \cite{Kawamura_benchmark}. The $T_c$ values trend better with $\lambda$ than $N(E_\mathrm{F})$. This agrees with the cases for Nb$_2$C \cite{Sevik_2023, Sevik_corr} and for superconducting bare MXenes \cite{Bekaert_MXene}. We also note that $T_c$ does not trend with lattice parameters (see Table \textcolor{blue}{S1} of \textcolor{blue}{Supplemental Material} for crystal structure data \cite{Supplemental_Material}).

\begin{figure}[t]
    \centering
    \includegraphics[width = 1\linewidth]{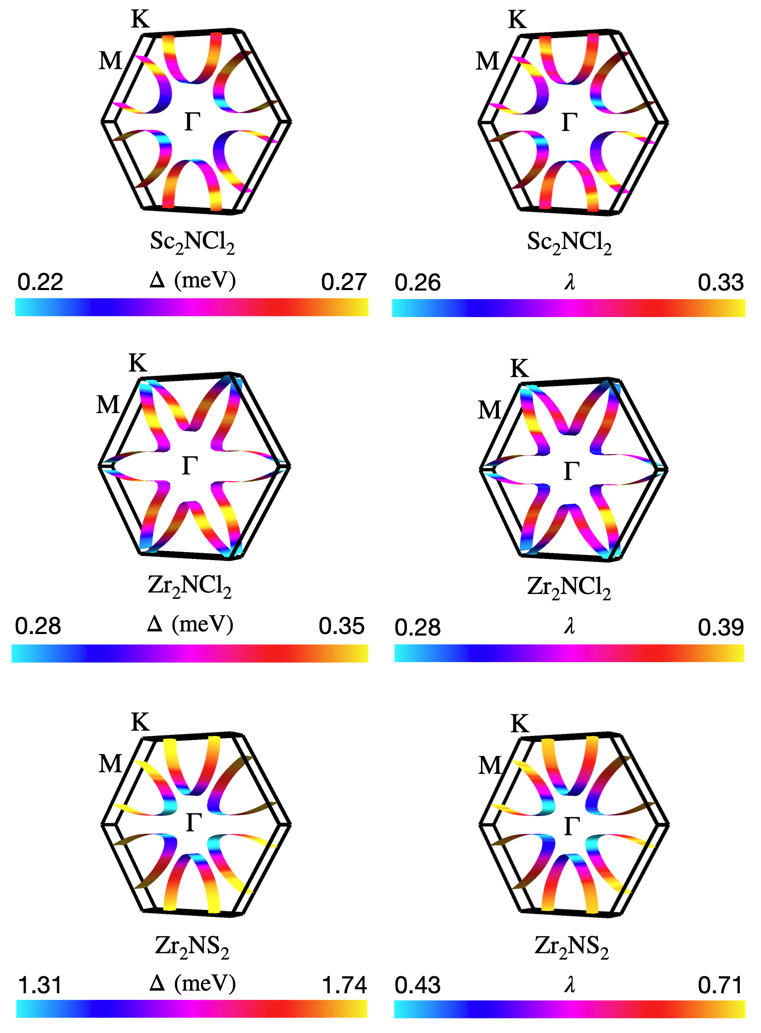}
    \caption{The electron-phonon coupling $\lambda$ and superconducting gap $\Delta$ profiles at the Fermi surface across the Brillouin zone for Sc$_2$NCl$_2$, Zr$_2$NCl$_2$, and Zr$_2$NS$_2$ at 1 K.}
    \label{fig:gap_lambda}
\end{figure}

\begin{figure*}[t]
    \centering
    \includegraphics[width = 1\linewidth]{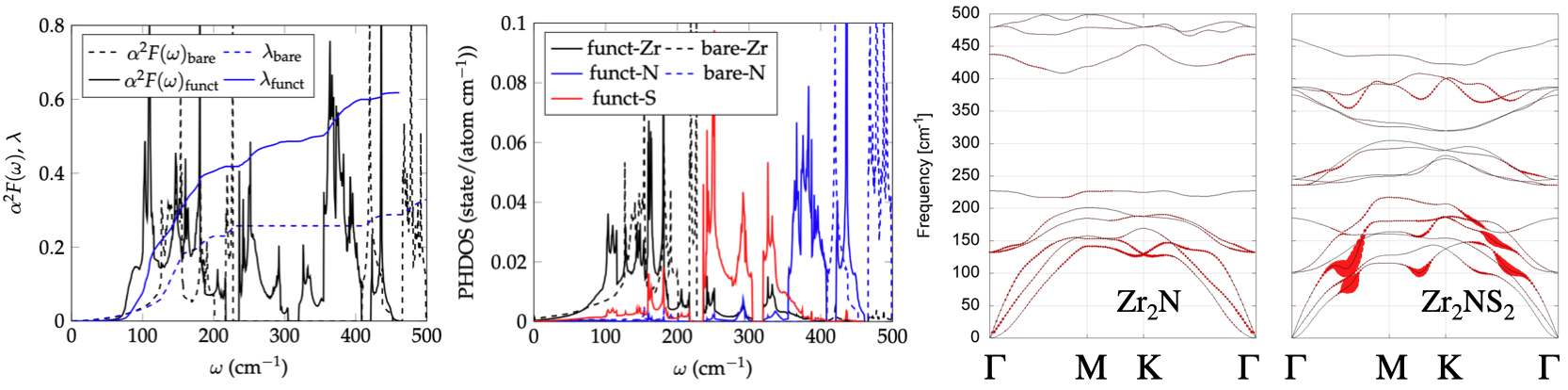}
    \caption{\textbf{Left}: Eliashberg spectral function $\alpha^2F(\omega)$, electron-phonon coupling $\lambda$, and partial phonon density of states (PHDOS) for the bare (dashed lines) and S-functionalized Zr$_2$N MXenes (solid lines). \textbf{Right}: Comparison of the phonon bandstructures (solid black lines). The red-colored circles are scaled equally for both compounds to their phonon-dependent electron-phonon coupling constants $\lambda_\mathbf{q\nu}$ across the Brillouin zone.}
    \label{fig:dispersion_Zr2NS2}
\end{figure*}

The profiles for $\lambda$ and SCDFT gap $\Delta$ at the Fermi surface across the Brillouin zone (BZ) are shown in Figure \ref{fig:gap_lambda}. In all compounds, we notice that the value distributions of $\lambda$ (color maps of Figure \ref{fig:gap_lambda}) resembles those of $\Delta$, indicating their correlation. We also plot $\Delta$ as a function of temperature $T$ in Figure \textcolor{blue}{S1} of \textcolor{blue}{Supplemental Material} \cite{Supplemental_Material}. The $(\Delta,T)$ curves have similar shapes for the studied compounds. 

\subsubsection{Understanding $\lambda$ enhancement in functionalized Zr$_2$N}

We compare Zr$_2$N and Zr$_2$NS$_2$ systems to illustrate the effects of functionalization in improving $\lambda$. Let us first consider the following expression for $\lambda$ \cite{Bekaert_MXene_H, QE1,QE2}:
\begin{equation}\label{eq-lambda1}
    \lambda = 2 \int \frac{\alpha^2F(\omega)}{\omega}d\omega
\end{equation}
and the isotropic Eliashberg spectral function,
\begin{equation}\label{eq-lambda2}
    \alpha^2F(\omega) = \frac{1}{N(E_F)} \sum_{\nu,\mathbf{k,q}} |g^\nu_\mathbf{k,k+q}|^2 \delta(\omega-\omega_\mathbf{q}^\nu) \delta(\varepsilon_\mathbf{k}) \delta(\varepsilon_\mathbf{k+q})
\end{equation}
with the electron-phonon matrix elements $(g^\nu_\mathbf{k,k+q})$, phonon $(\omega^\nu_\mathbf{q}$, mode index $\nu)$ and electron $(\varepsilon_\mathbf{k}$, measured from Fermi level$)$ bandstructures obtained from standard DFT+DFPT routines. Based on Eqs. \eqref{eq-lambda1}-\eqref{eq-lambda2}, we focus on low-frequency phonons to analyze the changes in $\lambda$.

The isotropic Eliashberg spectral function and its corresponding $\lambda$ for Zr$_2$N and Zr$_2$NS$_2$ are compared in Figure \ref{fig:dispersion_Zr2NS2}. We remark that these plots are computed separately from the SCDFT calculations in Table \ref{tbl:all}. In particular, the interpolation method \cite{wierzbowska2006,QE1,QE2} which enables efficient electron-phonon calculations using unshifted $\mathbf{q}$-point grids is used instead of the tetrahedron method at the DFPT step. A slight difference in the $\lambda$ values relative to Table \ref{tbl:all} may hence exist, but the physical insights should remain unaffected. We use a broadening parameter of 0.033 Ry, similar to Ref. \cite{wierzbowska2006}. This choice yields $\lambda$ value for Zr$_2$NS$_2$ of approximately 0.62, which are quite close to the SCDFT-obtained value of 0.6 in Table \ref{tbl:all}. Most of the $\lambda$ values are accumulated at low frequencies, with $\lambda$ of Zr$_2$NS$_2$ having reached 0.4 (2/3 of the total $\lambda$) under the frequencies of 200 cm$^{-1}$, confirming the importance of low-frequency phonons. In this regime, $\alpha^2F(\omega)$ has more spectral weight for Zr$_2$NS$_2$ than the bare compound. The $\lambda$ further increases to approximately 0.6 as the frequency reaches $\omega = 400$ cm$^{-1}$, in which a significant amount $\alpha^2F(\omega)$ spectral weight is present for the functionalized compound.

The aforementioned changes can be explained by comparing the partial phonon density of states (PHDOS) and the phonon dispersion shown in Figure \ref{fig:dispersion_Zr2NS2}. First, the S-contributed phonon states can be identified around $\omega \approx 250-350$ cm$^{-1}$. We note the Zr- and N-dominant phonon states for $\omega < 200$ cm$^{-1}$ and $\omega \approx 350-400$ cm$^{-1}$ having shifted from higher frequencies compared to the bare compound. These phonon softening can also be observed from the phonon dispersion plots. The PHDOS peaks may be explained by additional flat band sections and optimum points in the phonon dispersion, such as the one located just above $\omega \approx 100$ cm$^{-1}$ between M and K point. We also visualize the estimated phonon-dependent electron-phonon coupling constant $\lambda_\mathbf{q\nu} = \frac{1}{\pi N(E_\mathrm{F})} \frac{\gamma_{\mathbf{q\nu}}} {\omega^2_\mathbf{q\nu}}$ by scaling them to the red-colored marker size drawn on the dispersion lines. Unlike in the bare compound, the value of $\lambda_\mathbf{q\nu}$ is larger at parts of the dispersion with local minima at low frequencies. This finding is understandable considering the shape of the electronic Fermi surface of Zr$_2$NS$_2$ and also that local minimas in the phonon dispersion often translate to large phonon density of states.

\begin{figure}[h]
    \centering
    \includegraphics[width = 1\linewidth]{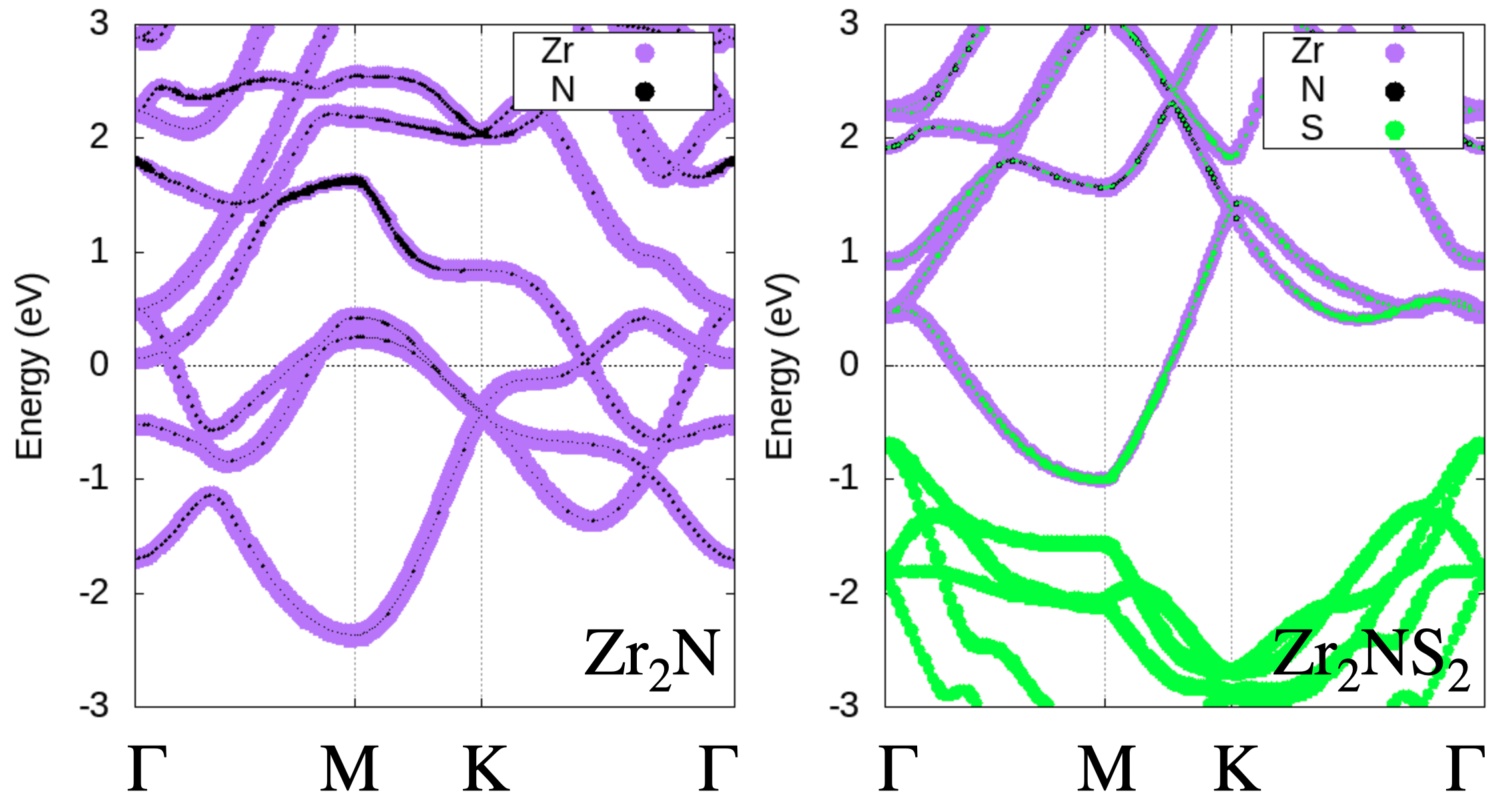}
    \caption{Projected electronic bandstructure of Zr$_2$N and Zr$_2$NS$_2$.  Purple, black, and green circles denote Zr, N, and S contributions, respectively.}
    \label{fig:projected_band}
\end{figure}

Next, the projected electronic bandstructures of Zr$_2$N and Zr$_2$NS$_2$ are shown in Figure \ref{fig:projected_band}. The Fermi level shifts to a lower energy, which suggests hole doping upon functionalization. The flat sections at $\Gamma$ and near K point in the bare compound are pushed above $E_\mathrm{F}$ in Zr$_2$NS$_2$, which can explain its reduced $N(E_F)$ value from 3.7 state/eV to 1.25 state/eV. There are significant contributions of S atoms at the Fermi surface between M and K points. As we do find between these points also a modified phonon dispersion (c.f., Figure \ref{fig:dispersion_Zr2NS2}) with high $\lambda$ and $\Delta$ values (c.f., Figure \ref{fig:gap_lambda}), hybridization between Zr and S bands might be a factor that contributes to such effects.

\subsubsection{Analyzing the $T_c$ difference between Sc$_2$NCl$_2$ and Zr$_2$NS$_2$}

From Table \ref{tbl:all}, we note that there is a significant difference between the $T_c$ values of Sc$_2$NCl$_2$ ($1.79$ K) and Zr$_2$NS$_2$ ($9.48$ K). This finding is interesting as their general shape of the Fermi surface is quite similar (see Figure \ref{fig:gap_lambda}). Furthermore, Sc$_2$NCl$_2$ has higher density of states at Fermi level as well as a smaller atomic mass than Zr$_2$NS$_2$. These traits may suggest that Sc$_2$NCl$_2$ should be having higher $T_c$ values than Zr$_2$NS$_2$ instead. Hence, we believe it is worthwhile to discuss more about the low $T_c$ of Sc$_2$NCl$_2$ predicted by our calculations here.

In Figure \ref{fig:gap_lambda}, we can read the superconducting gap and the electron-phonon coupling profile across the Fermi surface. Although there is a similarity in the Fermi surface shape, the superconducting gap size and its anisotropy are larger for Zr$_2$NS$_2$ relative to Sc$_2$NCl$_2$. These properties correlate with the electron-phonon coupling profile and are likely due to the interactions between the transition metal and functional group atoms near the Fermi level. As discussed in the previous subsection, the S-phonon states seem to shift the Zr and N phonon states in Zr$_2$NS$_2$ quite significantly toward lower frequencies. This was not quite the case for Sc$_2$NCl$_2$, where only a small fraction of the phonon states is found below $\omega = 150$ cm$^{-1}$ (see Figure \textcolor{blue}{S2} of \textcolor{blue}{Supplemental Material} \cite{Supplemental_Material}).

\begin{figure}[h]
    \centering
    \includegraphics[width = 1\linewidth]{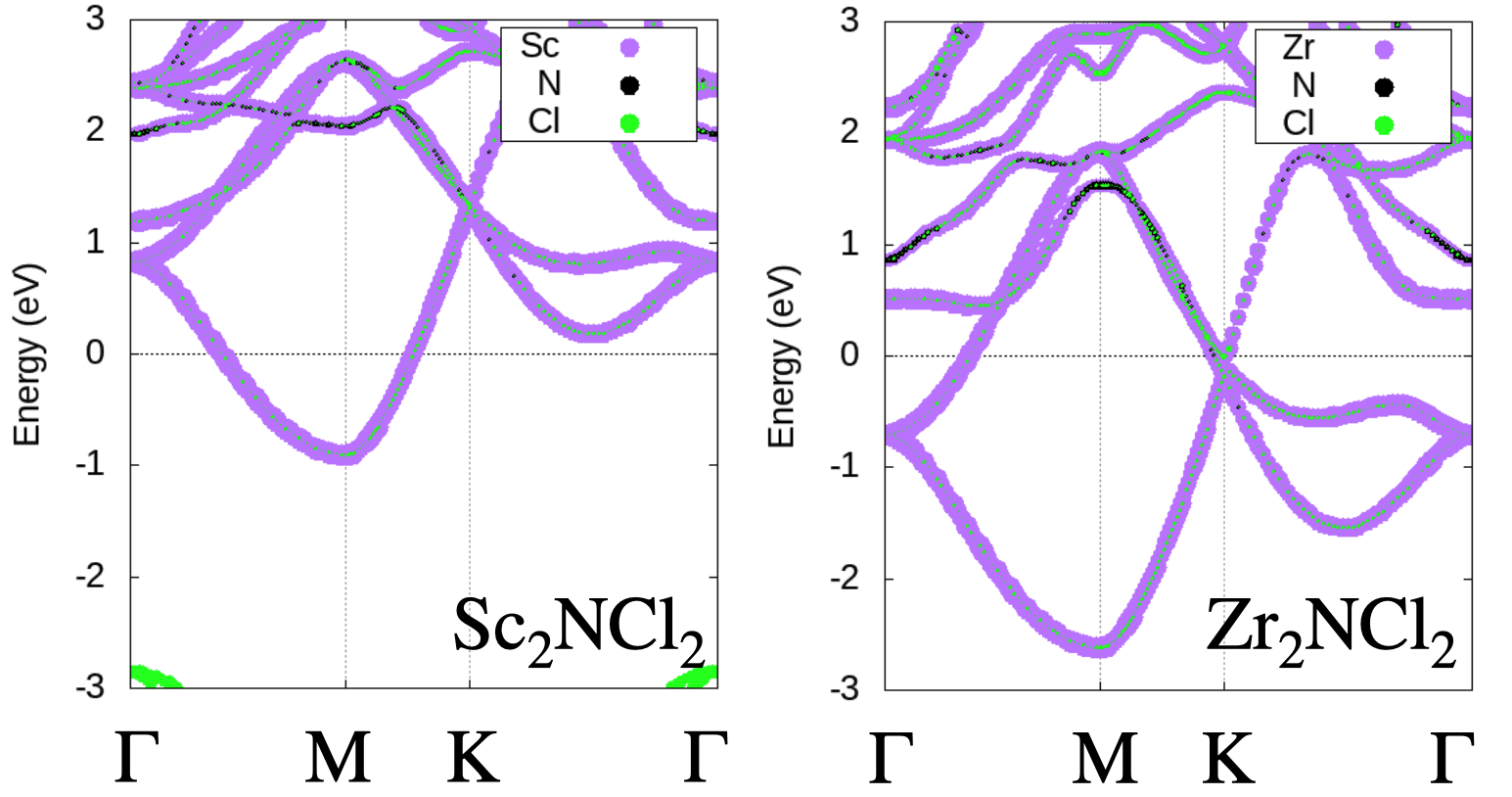}
    \caption{Projected electronic bandstructure of Sc$_2$NCl$_2$ and Zr$_2$NCl$_2$.  Purple, black, and green circles denote Sc/Zr, N, and Cl contributions, respectively.}
    \label{fig:projected_band_Cl}
\end{figure}

The projected electronic band structure for Sc$_2$NCl$_2$ is shown in Figure \ref{fig:projected_band_Cl}. We compare the band structure contributions from functional group atoms at the Fermi level between Sc$_2$NCl$_2$ and Zr$_2$NS$_2$. While S has a significant presence in the valence band of Zr$_2$NS$_2$ (see Figure \ref{fig:projected_band}), the presence of Cl in the valence band of Sc$_2$NCl$_2$ is more subtle. As suggested in the previous section, we believe this difference may be responsible for the distinct electron-phonon coupling strength between the two compounds which ultimately affects the superconducting gap anisotropy as well as the $T_c$ value.

On the other hand, we may also compare the Cl-functionalized compounds in which their Cl presence at the valence band is subtle: Sc$_2$NCl$_2$ and Zr$_2$NCl$_2$ (see Figure \ref{fig:projected_band_Cl}). From Table \ref{tbl:all}, the $T_c$ and $\lambda$ values are quite comparable to each other. The profiles in Figure \ref{fig:gap_lambda} are also similar in both magnitude and anisotropy. The slightly lower $T_c$ for Sc$_2$NCl$_2$ may be explained by the stronger electron-electron interaction parametrized by $\mu_C$. Together with the previous comparison to Zr$_2$NS$_2$, this observation suggests that the $T_c$ difference between these compounds might be indicated by the extent of each functional group's presence at the Fermi level.

Usually, the $T_c$ value may be increased by improving the electronic density of states $N(E_F)$ or the electron-phonon coupling. Because the bandstructure of bare MXenes may consist of several flat sections (see Figure \ref{fig:projected_band}), they can have higher value of $N(E_F)$ than their functionalized counterparts. However, many bare MXenes are not expected to superconduct due to their small $\lambda$ values despite having large density of states \cite{Bekaert_MXene}. Hence, we believe $\lambda$ has a bigger role than $N(E_F)$ in determining $T_c$ for these materials. The lower $T_c$ value of Sc$_2$NCl$_2$ despite its higher $N(E_F)$ relative to Zr$_2$NS$_2$ can therefore be explained in this framework.

\subsubsection[Strain-induced transition temperature enhancement in Cl-functionalized zirconium-nitride MXenes]{Strain-induced $T_c$ enhancement}
We move to discuss the effects of applied tensile strain. As we could not afford to perform a comprehensive strain optimization study due to limited computational resources, we demonstrate the possibility of strain-induced $T_c$ enhancement in a simple exercise for functionalized Zr$_2$N MXenes as follows. 

We begin by computing a selected compound (e.g. Zr$_2$NS$_2$) with an arbitrary strain value (e.g. 6\%) and compare its $N(E_F)$ value against the unstrained system. We then proceed to calculate its phonon dispersion if there is an increase in $N(E_F)$ value. The remaining SCDFT procedure are carried out only if no imaginary phonon frequencies are found. Otherwise, we reduce the strain level and repeat this cycle until these conditions are satisfied. For Zr$_2$NS$_2$ and Zr$_2$NCl$_2$, we arrive at the suitable strain level of 2\% and 6\% respectively.

\begin{table}[h]
  \caption{The superconducting transition temperature $T_c$ (K), electron-phonon coupling constant $\lambda$, electron density of states at Fermi level $N(E_\mathrm{F})$ (state/eV), and averaged Coulomb interaction $\mu_C$ for strained Zr$_2$NCl$_2$ and Zr$_2$NS$_2$.}
  \label{tbl:strain}
\begin{ruledtabular}
\begin{tabular}{lcccccc}
    Compounds & $T_c$   & $\lambda$ & $N(E_\mathrm{F})$  & $\mu_C$ \\
    \hline
   Zr$_2$NCl$_2$ (6\% strain) & 4.22 & 0.46 & 1.45 & 0.23 \\
   Zr$_2$NS$_2$ (2\% strain) & 11.58 & 0.76 & 1.37 & 0.20 \\
\end{tabular}
\end{ruledtabular}
\end{table} 

\begin{figure}[h]
    \centering
    \includegraphics[width = 1\linewidth]{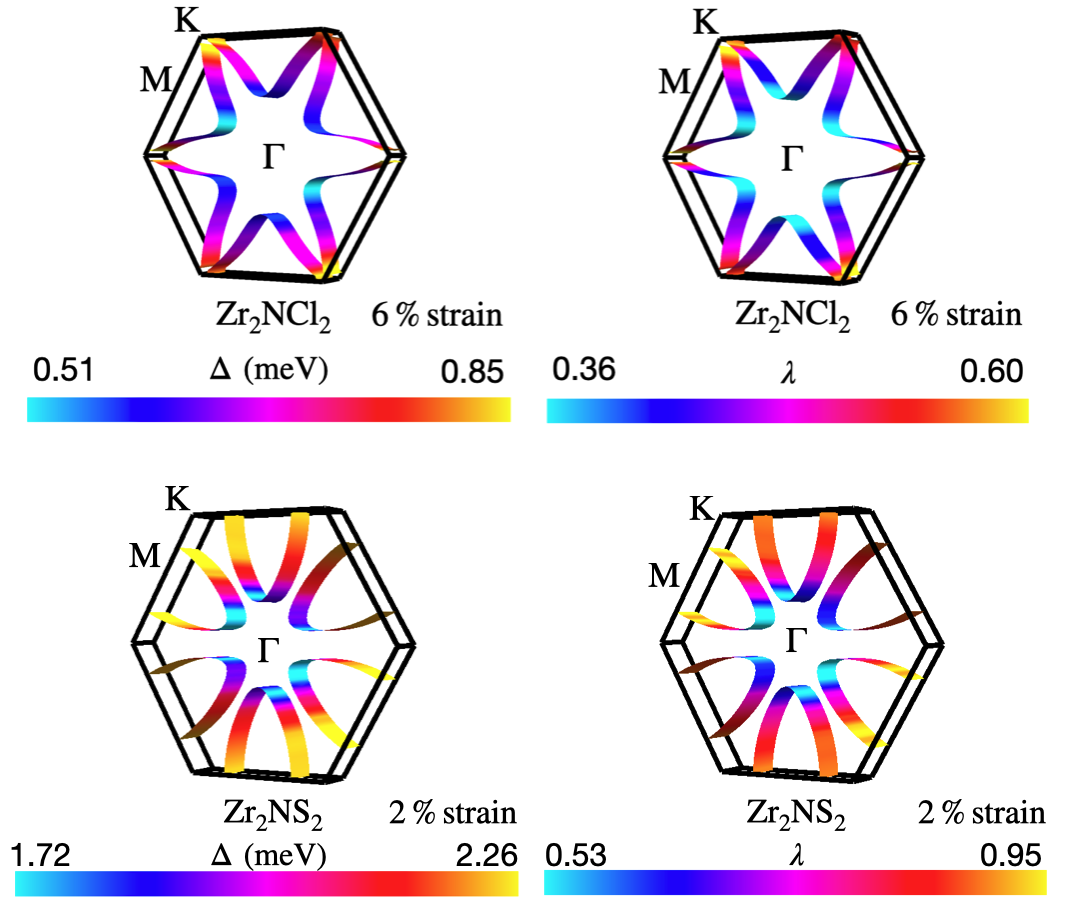}
    \caption{The superconducting gap $\Delta$ and the electron-phonon coupling $\lambda$ profiles at the Fermi surface across the Brillouin zone for Zr$_2$NCl$_2$ with 6\% strain, and Zr$_2$NS$_2$ with 2\% strain at 1 K.}
    \label{fig:gap_lambda_strain}
\end{figure}

We list the $T_c,\lambda$, $N(E_F)$, and $\mu_C$ values on the selected strained compounds in Table \ref{tbl:strain}, which are higher compared to their unstrained counterparts. The anisotropy in $\Delta$ and $\lambda$ are also enhanced, in particular for Zr$_2$NCl$_2$ with high values near the K point (Figure \ref{fig:gap_lambda_strain}). 

\begin{figure}[h]
    \centering
    \includegraphics[width = 1\linewidth]{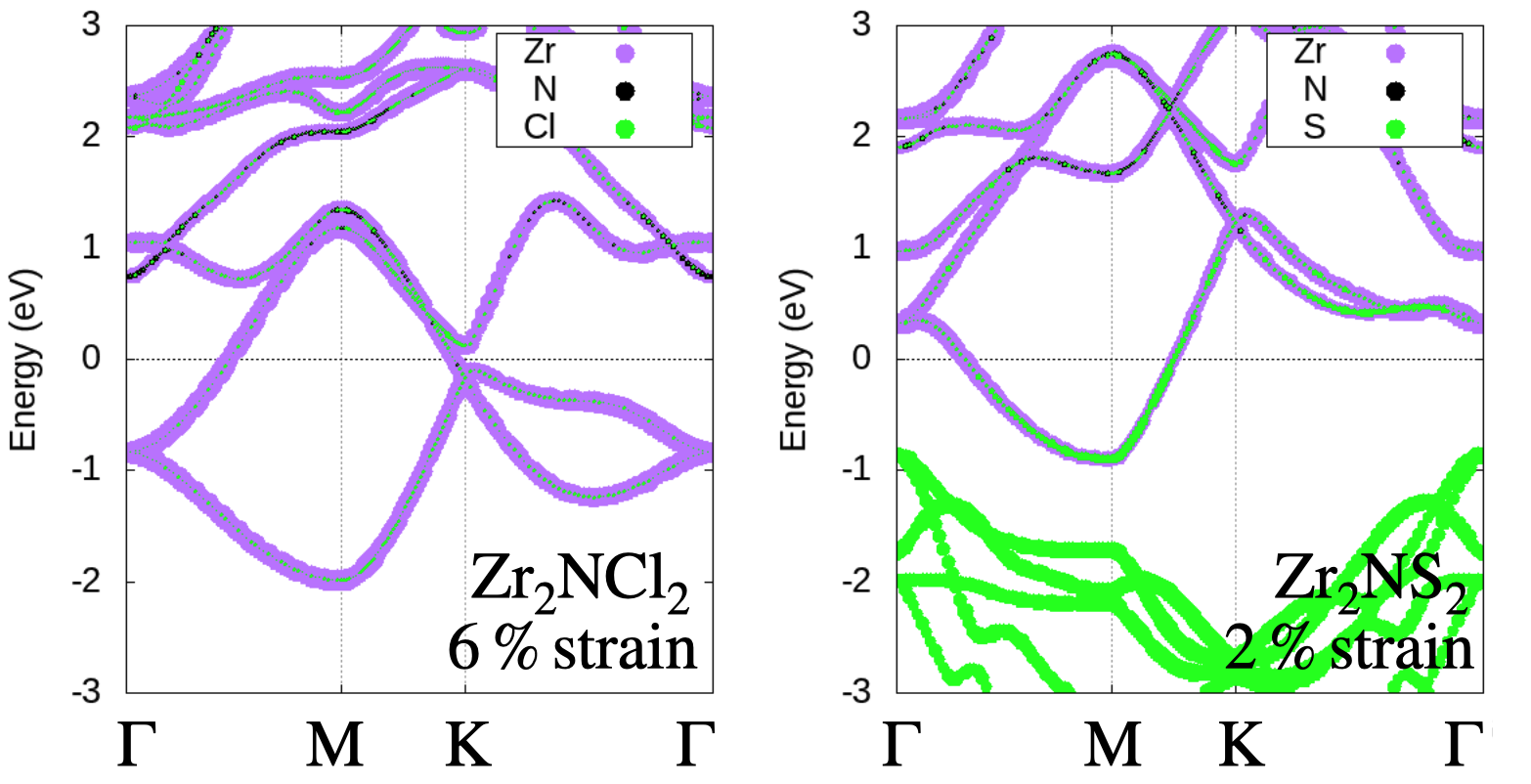}
    \caption{Projected electronic bandstructures for Zr$_2$NCl$_2$ (6\% strain) and Zr$_2$NS$_2$ (2\% strain). Purple, black, and green circles denote Zr, N, and Cl/S contributions, respectively.}
    \label{fig:projected_band_strain}
\end{figure}

We plot the projected electronic bandstructure in Figure \ref{fig:projected_band_strain}. The Zr$_2$NS$_2$ bandstructure does not show major changes for the valence band profile, which is not surprising as the applied strain is small. However, the bandstructure profile changes for Zr$_2$NCl$_2$ near $E_F$ around the K point, as it shows a flatter profile consistent with its increased $N(E_F)$. The functional group atoms still contribute to the valence band, with the contributions from S being larger than Cl for their respective compounds. Similar to our previous analysis, we note a shift of Zr phonon dispersion toward lower frequencies. (Figures \textcolor{blue}{S3 and S4} of \textcolor{blue}{Supplemental Material} \cite{Supplemental_Material}). The enhanced $\lambda$ values may thus be considered as a result from these changes in electronic and phononic bandstructure.

\subsection{Spin-fluctuation effects}

The effects of including ferromagnetic spin fluctuations (SF) \cite{essenberger_PhysRevB.90.214504, tsutsumi_PhysRevB.102.214515} as additional terms in the SCDFT exchange-correlation kernels are briefly discussed here.  $K^\mathrm{SF}_{n\mathbf{k}n'\mathbf{k'}}$, for instance, can be expressed as \cite{Kawamura_benchmark}: 
\begin{equation}
    K^\mathrm{SF}_{n\mathbf{k}n'\mathbf{k'}} (\xi,\xi') = \frac{2}{\pi} \int^\infty_0 d\omega \frac{|\xi|+|\xi'|}{(|\xi|+|\xi'|)^2+\omega^2} \Lambda^\mathrm{SF}_{n\mathbf{k}n'\mathbf{k'}}(i\omega)
\end{equation}
where $\xi$ again represents the Kohn-Sham eigenvalue and $\Lambda^\mathrm{SF}_{n\mathbf{k}n'\mathbf{k'}}$ is a summation of SF-mediated interactions:
\begin{align}
    \Lambda^\mathrm{SF}_{n\mathbf{k}n'\mathbf{k'}}(i\omega) = \sum_{\alpha = x,y,z} &\iint d^3r ~d^3r' \Lambda^\mathrm{SF}_{\alpha\alpha}(\mathbf{r,r'},i\omega) \nonumber \\
    &\times \rho^{(\alpha)}_{n\mathbf{k}n'\mathbf{k'}} (\mathbf{r}) \rho^{(\alpha)*}_{n\mathbf{k}n'\mathbf{k'}} (\mathbf{r'})
\end{align}
with the expressions involving Kohn-Sham orbitals $\varphi_{n\mathbf{k}\sigma}$,
\begin{align}
    \rho^{(x)}_{n\mathbf{k}n'\mathbf{k'}} (\mathbf{r}) &= \sum_{\sigma = \uparrow,\downarrow} \varphi^{*}_{n\mathbf{k}\sigma} (\mathbf{r}) \varphi_{n'\mathbf{k'}-\sigma}(\mathbf{r}) \\
    \rho^{(y)}_{n\mathbf{k}n'\mathbf{k'}} (\mathbf{r}) &= \sum_{\sigma = \uparrow,\downarrow} \sigma \varphi^{*}_{n\mathbf{k}\sigma} (\mathbf{r}) \varphi_{n'\mathbf{k'}-\sigma}(\mathbf{r})  \\
    \rho^{(z)}_{n\mathbf{k}n'\mathbf{k'}} (\mathbf{r}) &= \sum_{\sigma = \uparrow,\downarrow} \sigma \varphi^{*}_{n\mathbf{k}\sigma} (\mathbf{r}) \varphi_{n'\mathbf{k'}\sigma}(\mathbf{r}) 
\end{align}
and 
\begin{align}
    \Lambda^\mathrm{SF}_{\alpha\alpha}(\mathbf{r,r'},i\omega) = &- \iint d^3 r_1 d^3 r_2 I^{\alpha \alpha}_{XC} (\mathbf{r,r_1}) \nonumber \\
    &\times \Pi^{\alpha\alpha} (\mathbf{r}_1, \mathbf{r}_2, i\omega) I^{\alpha \alpha}_{XC} (\mathbf{r}_2, \mathbf{r'})
\end{align}
The spin susceptibilities of the Kohn-Sham ($\Pi^{\alpha\alpha}_\mathrm{KS}$) and interacting systems ($\Pi^{\alpha\alpha}$) are respectively given by \cite{Kawamura_benchmark},
\begin{align}
    \Pi^{\alpha\alpha}_\mathrm{KS} (\mathbf{r}, \mathbf{r'}, i\omega) = \sum_{\mathbf{kk'}nn'}  &\frac{\theta(-\xi_{n\mathbf{k}})-\theta(-\xi_{n'\mathbf{k'}})}{\xi_{n\mathbf{k}}-\xi_{n'\mathbf{k'}}+i\omega} \nonumber \\ 
    &\times \rho^{(\alpha)}_{n\mathbf{k}n'\mathbf{k'}} (\mathbf{r}) \rho^{(\alpha)*}_{n\mathbf{k}n'\mathbf{k'}} (\mathbf{r'})
\end{align}
\begin{align}
    \Pi^{\alpha\alpha} (\mathbf{r}, \mathbf{r'}, i\omega) &= \Pi^{\alpha\alpha}_\mathrm{KS} (\mathbf{r}, \mathbf{r'}, i\omega)  \nonumber \\ 
    &+ \iint d^3 r_1 d^3 r_2 \Big\{\Pi^{\alpha\alpha} (\mathbf{r}, \mathbf{r_1}, i\omega) \nonumber \\
    &\times I^{\alpha\alpha}_{XC}(\mathbf{r_1,r_2}) \Pi^{\alpha\alpha}_\mathrm{KS} (\mathbf{r_2}, \mathbf{r'}, i\omega) \Big\}
\end{align}
while the spin-spin interaction comes from the term,
\begin{equation}
    I^{\alpha\alpha}_{XC}(\mathbf{r,r'}) \equiv \frac{\delta^2 E_{XC}}{\delta m_\alpha (\mathbf{r}) \delta m_\alpha (\mathbf{r'})}
\end{equation}
which is the second-order functional derivative of the exchange-correlation energy with respect to the spin density along the $\alpha$ direction, $m_\alpha$ \cite{Kawamura_benchmark}.

Following Refs. \cite{tsutsumi_PhysRevB.102.214515, Berk_PhysRevLett.17.433},  SF effects may hinder the phonon-mediated pairing of electrons with opposite spins and effectively reduce $T_c$. Despite its importance, calculating the spin susceptibilities accurately near Fermi surfaces can be challenging, as the factor $\frac{\theta(-\xi_{n\mathbf{k}})-\theta(-\xi_{n'\mathbf{k'}})}{\xi_{n\mathbf{k}}-\xi_{n'\mathbf{k'}}+i\omega}$ may vary rapidly. Dense grids with stringent parameters should ideally be used in SF calculations, which significantly increase the numerical costs \cite{Kawamura_benchmark}. Because of our limited computing resources, we focus on simply demonstrating the qualitative side of SF effects, i.e., $T_c$ reduction. Using the same computational parameters as the non-SF case, we show in Table \ref{tbl:SF} the SF-corrected transition temperature $T^\mathrm{SF}_c$ (K), the Fermi surface average of SF kernel $\mu_S$ (c.f., Eq. \eqref{mu_C}):
\begin{equation}
    \mu_S = N(E_\mathrm{F}) \left\langle K^\mathrm{SF}_{n\mathbf{k}n'\mathbf{k}'} \right\rangle
\end{equation}
and the effective electron-electron interaction $\mu_C+\mu_S$ for our compounds.

\begin{table}[h]
  \caption{The computed SF-corrected superconducting transition temperature $T^\mathrm{SF}_c$ (K), averaged SF kernel $\mu_S$ and effective electron-electron interaction $\mu_C+\mu_S$.}
  \label{tbl:SF}
\begin{ruledtabular}
\begin{tabular}{lccc}
     Compounds & $T^\mathrm{SF}_c$ & $\mu_S$ & $\mu_C+\mu_S$\\
    \hline
  Sc$_2$NCl$_2$  & 0.89  & 0.06 & 0.33   \\
  Zr$_2$NCl$_2$  & 1.37  & 0.04 & 0.24    \\
  Zr$_2$NS$_2$ & 8.14  & 0.03 & 0.22    \\
\end{tabular}
\end{ruledtabular}
\end{table}

The SF inclusion does not affect the calculation of $\lambda$ or $N(E_F)$. However, SF leads to reduced $T^\mathrm{SF}_c$ values for all compounds. This reduction is likely due to the increased effective electron-electron interaction $\mu_C +\mu_S$. Although the $T^\mathrm{SF}_c$ values of Zr$_2$NCl$_2$ and Sc$_2$NCl$_2$ are reduced below 2 K, the $T^\mathrm{SF}_c$ for Zr$_2$NS$_2$ remains above 8 K. Hence, this result adds confidence in Zr$_2$NS$_2$ as a potential superconducting candidate.

In the prior SCDFT study for elemental superconductors, superconductivity for elemental Sc is completely suppressed $(T_c: 2.7 \rightarrow 0 ~\mathrm{K})$ by including SF effects \cite{Kawamura_benchmark}. This suppression is likely due to Sc's high $\mu_S$ value (0.97), which in turn may be caused by the localized $3d$ orbital and its high density of states (2.01 states$/$eV) \cite{Kawamura_benchmark}. On the other hand, our compounds in Table \ref{tbl:SF} have smaller $\mu_S$ values which cause their superconductivity not to be fully suppressed. This difference in $\mu_S$ values may be due to several interrelated factors as follows.

First, our MXene compounds have smaller density of states at the Fermi level [$N(E_F)<1.5$ state$/$eV, see Table \ref{tbl:all}]. Next, the electron configuration of Zr has valence electrons in the $4d$ orbital. The delocalization of electronic orbitals increases with the principal quantum number $3d \rightarrow 4d$, and SF effects should be weaker in less localized orbitals \cite{tsutsumi_PhysRevB.102.214515, Kawamura_benchmark}. Furthermore, the MXene valence band is also contributed by N, Cl, and S atoms. The valence configurations of these atoms are made of $s$ and $p$ orbitals, which are less localized than the $3d$ orbital of elemental Sc. We also observe that these compounds do not have many flat sections in their electronic bandstructure at the Fermi level. Since flat band sections indicate strong localization with typically high $N(E_F)$, their absence is in agreement for a more delocalized picture that leads to a small SF parameter $\mu_S$.

Nevertheless, we re-emphasize that SF calculations are more sensitive to computational parameters and thus often requires higher cutoff energies to be precisely computed than non-SF calculations. Unfortunately, this complexity may cause the computational costs to become prohibitive, even for simple elemental superconductors. More resources are required for 2D compounds as vacuum must be included in the computational cell. As such, our SF study should be regarded only as a qualitative exercise. In the case of elemental Scandium, its $T_c$ is completely suppressed with SF effects. Thus, SF can qualitatively change the predicted property for a material from superconducting to non-superconducting. By performing our simple SF calculations, we simply wish to verify whether our compounds, especially Zr$_2$NS$_2$, remain good superconductor candidates even when SF effects are included. As reflected in Table \ref{tbl:SF}, this seems to be the case as Zr$_2$NS$_2$ remains with a sufficiently high $T^\mathrm{SF}_c$ which can be reasonably measured in experiments.

\subsection{Possible extensions}

The examples presented in this brief study are not intended to be exhaustive. Wider choices of functional groups or MXene compounds may be explored with the help of structure prediction algorithms (e.g., Ref. \cite{aek_2023}), and further refinements to the SCDFT calculations may be opted with added complexities. Inclusion of van der Waals corrections may also be pursued if a suitable exchange-correlation functional can be identified. MXene layers are reported to be only weakly coupled in prior studies \cite{Sevik_2023, Sevik_corr}. For Nb$_2$C systems, attempts \cite{Sevik_2023,Sevik_corr} to include van der Waals corrections based on Grimme's DFT-D2 \cite{DFT-D2}, DFT-D3 \cite{DFT-D3}, and Becke-Johnson method \cite{Becke_Johnson} were not successful. Careful treatment of van der Waals effects is a desirable extension in future studies.

\section{Conclusions \label{sect-conclusions}}
In this paper, we explore several new superconductor candidates in functionalized MXenes beyond the Nb$_2$C system. Zr$_2$NS$_2$, Zr$_2$NCl$_2$, and Sc$_2$NCl$_2$ are studied with density functional theory for superconductors (SCDFT). $T_c$ is computed without empirical parameters, hence differentiating this study from prior works \cite{Bekaert_MXene, Bekaert_MXene_H, Sevik_2023, Sevik_corr}. The $T_c$ is predicted to reach as high as 9.48 K for Zr$_2$NS$_2$, while further improvements are possible with applied strain. The $T_c$ trends with $\lambda$, whose profile across the Fermi surface resembles that of $\Delta$. The $\lambda$ enhancement is suggested to come from modified phonon dispersion and electronic bandstructure. Hybridization between the transition metal and functional group atoms may induce such modifications, as reflected from their contributions in the electronic bandstructure at the Fermi level. We encourage further research to explore more functionalized MXene compounds as superconductor candidates.

\section*{Author Contributions}
\textbf{Alpin N. Tatan}: Conceptualization, Methodology, Formal analysis, Investigation, Writing - original draft, Writing - review and editing. \textbf{Osamu Sugino}: Funding acquisition, Supervision, Writing - review and editing.

\section*{Conflicts of interest}
There are no conflicts to declare.

\section*{Data availability}
The data supporting this article have been included as part of the Supplemental Material \cite{Supplemental_Material}.
\begin{acknowledgments}
The calculations were performed with the facilities of the Supercomputer Center, the Institute for Solid State Physics, the University of Tokyo.
\end{acknowledgments}

\bibliography{references}

\end{document}